\documentclass[twocolumn,aps,prb,graphicx,showpacs]{revtex4}%
\usepackage{amsfonts}
\usepackage{amsmath}
\usepackage{amssymb}
\usepackage{graphicx}%
\begin{document}

\title{Towards a Classification of the Effects of Disorder on Materials Properties}
\author{A. J. Millis\\Department of Physics\\Columbia University\\538 W. 120th St., NY NY 10027}
\date{Oct 9, 2002}

\begin{abstract}
Many 'interesting; correlated electron materials exhibit an unusual
sensitivity of measured properties to external perturbations, and in
particular to imperfections in the sample being measured. It is argued that in
addition to its inconvenience, this sensitivity may indicated potentially
useful properties. A partial classification of causes of such sensitivity is given.

\end{abstract}
\maketitle

\section{Introduction}

In this brief note I discuss the effects of disorder on correlated electron
systems from a perhaps slightly unusual point of view. There is a large, if
mainly anecdotal, body of evidence indicating that 'interesting' correlated
electron materials are unusually sensitive to disorder. As an example, not
quite as frivolous as it may seem at first, I note that in the first few years
of investigation of any new class of 'interesting' materials, conferences are
dominated by debates concerning which samples exhibit 'intrinsic' behavior and
which measurments are meaningless because performed on 'bad' samples. The
ubiquity of these discussions suggests that there is a general and possibly
interesting phenomenon at work, in other words that one should pose the question:

\begin{center}
\textit{Why are so many 'interesting' materials so sensitive to (apparently
weak) disorder?}
\end{center}

The question may be of more than academic interest. A strong effect of
disorder on materials properties is simply one of many examples of
\textit{sensitivity of materials properties to perturbations. }This
sensitivity may be useful, or inconvenient, or both. A familiar example of a
useful sensitivity of properties to perturbations is provided by
semiconductors, where in an appropriate device geometry, the resistivity is
very sensitive to applied 'gate' voltages and this sensitivity is the basis of
the modern electronics industry. Along with this useful sensitivity comes an
inconvenient one: the properties of semiconductors are very sensitive to
defects, and indeed it took more than two decades of research to learn to
control the undesirable sensitivity so that the useful one could be exploited.
I suggest that one should view other examples of sensitivity of properties to
perturbations in the same light: that a sensitivity of measured properties to
sample imperfections may be an indication of some interesting, and potentially
useful, properties of the material, so that understanding and controlling this
phenomenon are important open issues in materials physics.

In what follows I\ present a first attempt to address these issues by
presenting a 'botany' of materials exhibiting unusal sensitivity to disorder
and a partial classification of mechanisms known to be operating in these
different cases. I pay special attention to the 'colossal' magnetoresistance
(CMR) materials, whose properties seem to indicate a qualitatively new
mechanism for sensitivity of properties to perturbations.

This paper is presented in the hope that the participants in the Williamsburg
Conference (and especially the muon-spin-rotation community, which has given
us so much beautiful information about inhomogeneous structures in correlated
materials) will find something of interest.

\section{Examples of Sensitivity of Properties to Perturbations}

\subsection{Overview}

The magnitude of the response induced in a material by a perturbation is
determined by comparing the size of the perturbation to some property of the
material. A large response implies some unusal system property. The cases so
far known to the author may be classified as follows:

\textit{(a)} \textit{Small parameter}: if some scale in the material is very
small, then it is reasonable to assume that even a weak external perturbation
may change the system properties dramatically. Two examples of this are the
semiconductor (where the small parameter is the electron density) and the
'Kondo disorder' picture of heavy fermion materials (where the small parameter
is the Kondo temperature)

\textit{(b) (Geometrical) Frustration: }in frustrated systems (for example
Ising spins on a triangular lattice) a thermodynamically large set of
constraints prevents the Hamiltonian from finding a 'natural' ground state,
leading (among other things) to a high degeneracy (or near-degeneracy) of low
lying states. Disorder, by lifting the frustration, may then rearrange this
large number of nearly degenerate states, thus qualitatively changing the
observed behavior.

(c) \textit{Proximity to second order phase transition (critical or quantum
critical point)}. In this case the divergent susceptibilities associated with
the critical point may couple to disorder, leading to large effects. This may
be thought of as a sub-case of \textit{(a)} with the inverse susceptibility
being the small parameter, but requires a separate discussion.

\textit{(d) Proximity to a first order transition.} \ Although first order
transitions are often considered to be uninteresting, it was shown many years
ago by Imry and Wortis that in appropriate circumstances disorder may have a
dramatic effect, leading to multiphase coexistence, percolative phenomena and
possible changes in the order of phase transitions.

\textit{(e) CMR materials: first order energy landscape, 'martensitic'
accomodation strain, and the importance of nonlinear response. }The recent
extensive study of the 'colossal' magnetoresistance' (CMR) materials has
revealed that the eponymous CMR is but one manifestation of a greatly enhanced
sensitivity of properties to perturbations, whose origins involve both a sort
of frustration and proximity to a first order phase transition.

In the rest of this section a more detailed discussion of examples
\textit{(a)-(e) }is presented.

\subsection{Small parameter}

As noted above, a very familiar example of sensitivity of properties to
perturbations is the semiconductor. Here the small parameter is the carrier
density, $n$. The low carrier density means that relatively modest changes in
external parameters such as a gate voltage can modulate this density and
therefore the conductivity of the device. It also means that the device is
very sensitive to disorder. In particular, it took many years of materials
work before the density of 'traps' (sites which capture an electron or hole)
could be reduced below the carrier density, so that intrinsic behavior could
be observed.

A different example the combination of weak disorder and a small energy scale
occurs in the 'Kondo disorder' model of non fermi liquid effects in heavy
fermion compounds. This picture was deduced by Bernal and co-workers from
their NMR data \cite{Bernal95} and was studied theoretically in some detail by
Dobrosavljevic and collaborators \cite{Dobrosavljevic97}. The physics at issue
is the 'non-fermi-liquid' (i.e. weakly divergent) magnetic susceptibility and
specific heat exhibited by a range of \ 'heavy electron' materials. The basic
physics is this: heavy electron metals involve local moments which are coupled
via an exchange coupling $J$ to an electronic conduction band characterized by
a fermi energy $E_{F}$. Both $E_{F}$ and $J$ are of a reasonable ($eV)$ order
of magnitude, although typically $J$ is smaller than $E_{F}$ by a reasonable
numerical factor. In this situation a lattice version of the Kondo effect
causes the local moments to dissolve into the conduction band, leading to a
'heavy fermi liquid' characterized by an energy scale conventionally denoted
$T_{K}$. For example the specific heat coefficient $\gamma=\lim_{T\rightarrow
0}\frac{C}{T}\sim1/T_{K}.$ The basic scale $T_{K}$ is given in terms of
electronic parameters by $T_{K}\sim E_{F}e^{-E_{F}/J}$. Thus a ratio $E_{F}/J$
which is not too much larger than unity can lead to an extremely small Kondo
temperature. \textit{\ }As suggested by Bernal et al \cite{Bernal95} and
convincingly demonstrated by Dobrosavljevic et al \cite{Dobrosavljevic97}
modest disorder can lead to a modest variation in $J$ which, because it is
amplified by the exponential factor can lead to very broad distribution of
$T_{K}$ and thus to a dramatic effect on low temperature properties. It is
important to note that this 'Kondo disorder' is not the only source of
non-fermi-liquid physics in heavy fermion materials. Novel single-impurity
physics and proximity to quantum critical points are believed also to play
some role (for reviews see e.g. \cite{Stewart01,Millis02a} but I believe that
the existence of the 'Kondo disorder' effect is not now in doubt.

\subsection{Frustration}

In 'frustrated' systems, constraints (often geometrical in nature) prevent the
system from achieving a gound state in which all interactions are satisfied.
For reviews see \cite{Moessner00,Ramirez96} . A classic example involves spins
located at the vertices of the 'pyrochlore' lattice shown in Fig. 1a and
interacting mainly by nearest-neighbor antiferromagnetic interactions. In this
situation, it is not possible to fully satisfy all bonds, but the many ways to
partially satisfy most of the interactions leads to a very large degeneracy of
low-lying states, leading e.g. to the large in the spin wave spectrum shown in
Fig. 1b. It is natural to expect that lattice distortions, either
spontaneously induced or caused by disorder, will lift the frustration and
therefore couple to the large density of low-lying states. In non-disordered
systems this leads to the interesting 'spin-Teller' effect introduced by
Yamada and Ueda \cite{Yamashita00} and by Tshcernyshev and co-workers
\cite{Tschernyshev02}. A similarly large effect from disorder may be anticipated.
\begin{figure}[ht]
\includegraphics[width=3.0in]{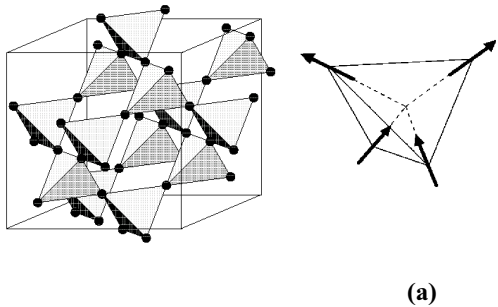}
\includegraphics[width=3.0in]{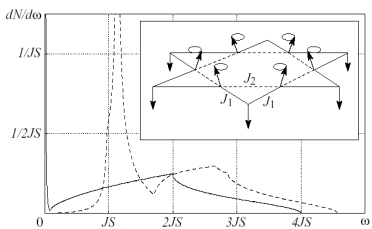}
\caption{Panel a: Left side: section of pyrochlore lattice, highlighting
tetrahedra of frustrated spins (from \cite{Moessner00}).
right side: example of spin arrangement partially satisfying
nearest neighbor antiferromagentic bonds (from \cite{Moessner00}).
Panel b: Theoretically expected spin wave spectrum in two cases: 
Solid line: undistorted lattice. Note the  sharp peak at $\omega =0$  
Dashed line: disorted lattice--note peak is shifted to much higher
energies by distortion. Inset: spin excitation giving
rise to sharp peak. From \cite{Tschernyshev02}. }
\end{figure}
\subsection{Proximity to a phase transition}

A second order phase transition involves a diverging length scale $\xi~$ and
diverging susceptibilities. Disorder which couples to these divergences can
have very large effects, which have been extensively studied. It is
conventional to represent the critical degrees of freedom by an order
parameter field $\phi$ and to model \ the static part of the energy via a
Landau Free energy for a system in an applied field $h_{0}$ which couples to
the uniform component of the order parameter:%
\begin{eqnarray}
F&=&\xi^{-2}\phi^{2}+\left(  \nabla\phi\right)  ^{2} +u\phi^{4} \nonumber \\
&+&\left(
h_{0}+h_{ran}(x)\right)  \phi(x)+m_{ran}(x)\phi^{2}(x)+....
\end{eqnarray}
It is useful to distinguish between \textit{random fields} (such as $h_{ran}$)
above, which couple linearly to the order parameter and thus locally 'tell it
which way to point' and a \textit{random mass} which couples to the square of
the order parameter, and may be thought of as changing the local value of the
transition temperature. Random field and random mass effects have been
extensively studied , and may in appropriate circumstances be very large. For
example, in a system with Ising symmetry at $h_{0}=0$ an arbitrarily weak
random field $h_{ran}$ will destroy the long range ordered state in spatial
dimension $d\leq2$ \cite{Imry75} while an arbitrarily weak 'random mass' will
be a relevant perturbation (thus changing e.g. the critical exponents
characterizing the phase transition) if the product of the pure system
correlation length exponent $\nu$ and the spatial dimensionality $d$ is less
than $2$ (for a discussion and references in the context of quantum ($T=0$)
phase transitions see e.g. \cite{Sachdev00}).

In both random field and random mass cases, when the randomness is important
(either because its strength is sufficiently large or because the
dimensionality is sufficiently low) the main effect is to produce 'droplets'
of one phase inside a region which is on average composed of the other, and to
change the character of the phase transition (if any) to a percolative one in
which, as a parameter is varied, droplets of one phase increase in size and
gradually connect, leading to a long ranged order.

Recently, Morr, Schmalian and the author \cite{Millis01a} studied perhaps the
simplest example of a 'droplet'--namely that nucleated by a single, localized
random mass defect. For a system sufficiently near a quantum critical point a
surprisingly 'universal' behavior of the space dependent droplet amplitude was
found (see Fig 2). Also, perhaps not very surprisingly, in three dimensional
systems line and plane defects (produced e.g. by screw dislocations or
stacking faults) are much more effective at nucleating droplets than are point
defects. It is tempting to speculate that such defects (which have been argued
to exist \cite{Aronson93} are responsible for some of the weak magnetism
observed \cite{Aeppli88} e.g. in heavy fermion materials such as $UPt_{3}$.

\begin{figure}[ht]
\includegraphics[width=3.0in,angle=270]{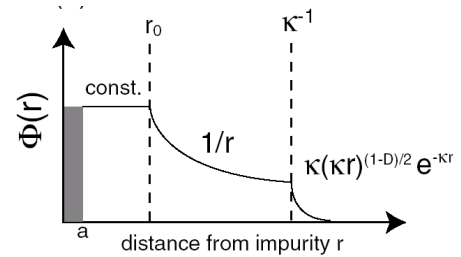}
\caption{Form of 'droplet' induced in metallic quantum critical
system by local defect. Here the 'core size'
$r_0$ depends on the defect strength and on the relative
dimensionality $D$ between the defect and the host material,
while $\kappa$ is the inverse correlation length of the host
system. From \cite{Millis01a}.}
\end{figure}

\subsubsection{Random fields and first order phase transitions}

The behavior of the Ising model in $d>1$ at very low temperatures and in an
applied field $h_{0}$ constitutes a very interesting example of this
phenomenon. If $h_{ran}=0$ then the model exhibits a first order phase
transition as $h_{0}$ is varied through $h_{0}=0$. (In the ground state, all
of the spins point \ parallel to $h_{0}$ and therefore change direction when
$h_{0}$ goes through $0$). In the presence of an arbitrarily weak random field
in $d=2$ or a sufficiently strong random field in $d=3$, the situation is
changed: for very large $h_{0}$ all the spins follow $h_{0}$, but as $h_{0}$
is reduced towards $0$, more and more of the spins follow the random field and
so domains of misoriented spins appear and as $h_{0}$ goes through zero the
mean polarization vanishes. Thus the random field has turned a first order
transition into a second order one. This observation was generalized by Imry
and Wortis \cite{Imry79}, who noted among other things that one could map a
generic system undergoing a first order phase transition onto the random-field
Ising model, by identifying the two different phases with the 'spin up' and
'spin down' phases of the Ising model and noting that \textit{any} randomness
coupling to the energy density would favor one phase more than the other and
would therefore behave as a random field.

Interestingly, while theoretical arguments strongly suggest that two and three
dimensional systems should behave very differently, experimental evidence
suggests that the behavior of a the classic random field system in $d=3$ is
quite similar to that expected theoretically for systems in $d=2$
\cite{Feng95}.

\subsubsection{One dimensional physics}

One dimensional (or quasi one dimensional) materials are in a certain sense
critical systems--the generic lwo temperature and low energy behavior involves
power law correlations with novel exponents, and it is not surprising that one
dimensional systems are unusually sensitive to disorder. A large theoretical
literature exists on this question which will not be summarized here. One very
interesting example however should be noted. The material $CuGeO_{3}$ is an
insulator . The $Cu$ site is magnetic (each $Cu$ has a $S=1/2$ magnetic
moment) and the magnetic couplings are such that the material should be
thought of as a collection of spin chains. and is most stongly coupled to two
but at low temperatures a non-magnetic state is formed, evidently because of a
spin-Peierls distortion which dimerizes the spin chains \cite{Hase93}.
Interestingly, extremely small $Zn$ doping was found to induce commensurate
antiferromagnetic order at a relatively high and doping independent
temperature \cite{Lussier95}, leading to interesting theoretical work
modelling impurities in spin chains. However, more recent experimental results
point to a different, non-magnetic origin for the results: the $Zn$ doping
acts as a 'random field' on the three dimensional spin-Peierls distortion, and
one consequency of the disruption of the spin Peierls order leads to the
magnetism \cite{Wang99}.

\section{'Colossal' magnetoresistance manganites}

The 'colossal' magnetoresistance (CMR) manganites, which are pseudocubic
perovskites of the form $\operatorname{Re}_{1-x}A_{x}MnO_{3}$ with
$\operatorname{Re}$ a rare earth and $A$ a divalent alkali \ (examples also
exist in the closely related Ruddlesden-Popper structures) offer a surprising
new paradigm for strong effects of weak disorder. These materials have been
known for a long time (and indeed were the subject of one of the first neutron
scattering investigations reported in the condensed matter physics literature
\cite{Wollan54}) and have been the subject of a great resurgence of interest
(for reviews see, e.g. \cite{CMRbook} since Jin and co-workers \cite{Jin94}
showed that in appropriatedly designed materials the magnetoresistance
(dependence of resistivity on magnetic field) could be made extremely large
('colossal'). Fig 3a shows an example of the eponymous magnetoresistance,from the early work of Schiffer et. al. \cite{Schiffer95}.

\begin{figure}[ht]
\includegraphics[width=2.9in,angle=270]{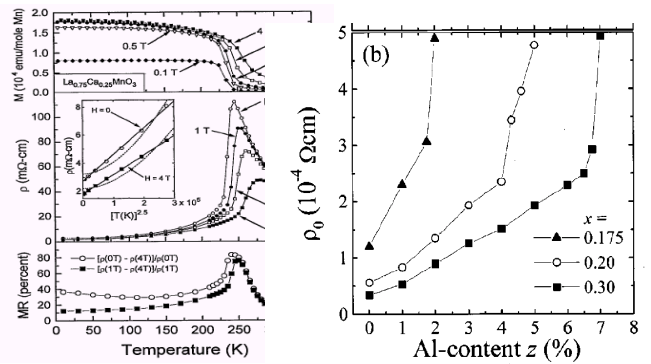}
\caption{Left side--middle panel: resistivity as function of temperature
for a 'CMR material, demonstrating large change of resistivity
with magnetic field. Upper panel: magentization as function of
temperature, demonstrating that large change in resistivity
is associated with magnetic-non-magnetic phase transition.
Data from \cite{Schiffer95}
Right side: Resistivity as function of aluminum doping, from \cite{Sawaki00}.
Note threshold behavior of resistivity.}
\end{figure}

The interesting point which has emerged from subsequent study is that the very
large magnetoresistance is but one example of \ a generically extreme
sensitivity of properties to perturbations. The right hand
side of Fig. 3 shows a very interesting
second example of this phenomenon: a sharp sensitivity of electrical
resistivity to changes in chemical composition (in this case, doping $Al$ into
the electrically active $Mn$ site) observed by Sawaki and co-workers
\cite{Sawaki00}. In the author's view the great conceptual importance of the
result shown in the panel is that it clearly demonstrates that the observed large
sensitivity is not due to a large linear response. The initial slope of the
$\rho$ vs $Al-concentration$ curve is consistent with conventional,
'unitarity-limit' expectations. The large effects only occur when the
concentration exceeds a small, material-dependent threshold, above which the
properties change dramatically. Note in particular the $x$-dependence of the
threshold concentration. In the undoped (no $Al$) material, $x\approx0.18$
marks the boundary between a larger $x$ ferromagnetic metal phase and a
smaller $x$ charge and orbitally ordered insulating phase. The systematic $x$
dependence of the $Al$ doping effects strongly suggest that, when the
($x-dependent)$ threshold is exceeded the effect of $Al$ substitution is
transform the material into the insulating phase.

The qualitative phenomena revealed by $Al$ doping, namely a linear response
which is not particularly large and an enormous nonlinear response
characterized by a low threshold for transforming the material into another
phase, seem to characterize all of the other enhanced responses in the CMR
materials. In particular, in all CMR materials, the very low field
magnetoresistance (the coefficient $\rho_{2}$ in the low field expansion
$\rho(B)=\rho_{0}+\frac{1}{2}\rho_{2}B^{2}$) is not especially large--in fact
it is rather smaller than the $\rho_{2}$ found in the 'GMR' multilayer devices
used in present-day magnetic read-heads. The truly large effects arise when
$B$ exceeds a low (order $1T$) threshold, above which material properties
change qualitatively from insulating to metallic, so the question
becomes: why is the threshold so low?

This phenomenon is not yet well understood, but multiphase coexistence due (in
some as yet mysterious way) clearly plays a
key role.  Important early work discussed electronically
driven phase segregation into two phases of differenct electronic density
\cite{Moreo99}, and helped introduce the concept of inhomogeneity
in the manganite context
but it seems to this author that  two crucial pieces of the physics and
materials science came from experiments. One is due to 
Fath and collaborators \cite{Fath99} and 
S-W. Cheong and co-workers \cite{Uehara99} who showed in thin
film (Fath) and bulk (thinned for TEM) (Cheong) materials that very large
domains of magnetic and non-magentic material can occur. 
cheong and collaborators extended this work, showing convincingly
that the 
CMR materials exhibiting the largest
magnetoresistance are tuned to be near a first order transition (in which a
putative completely un-disordered material would change ground state from
charge ordered insulator to ferromagnetic metal) and exhibit multiphase
coexistence (a term I prefer to phase separation), with large (up to
micron-scale) domains of ferromagnetic metal interleaved with similar size
non-ferromagnetic, charge and orbitally ordered insulator. The large size of
the domains guarantees that the phenomenon is not driven by charge
inhomogeneity. Subsequent experiments \cite{Podzorov00} showed in detail that
\ the 'colossal' effects were shown to arise from a percolation phenomenon, in
which as a parameter (e.g. field) was varied the volume fraction of conducting
material grew and eventually percolated. One may follow Imry and Wortis and
attempt to model this phenomenon in terms of the low-T behavior of the
random-field Ising model in a uniform applied field: (for a discussion and
references see e.g. the article of Burgy et. al. in this volume \cite{Burgy02}%
). However, the real-materials aspects of the energetics have not yet been
addressed. In particular, there is to the authors knowledge no understanding
of domain wall energies or stiffnesses.

The Cheong group also uncovered a second key aspect of the phenomenon
\cite{Podzorov01b}, namely an essential role of 'martensitic' accomodation
strain. The charge and orbitally ordered phase induces a long-ranged strain
which, in the absence of constraints, would cause a change in shape of the
material as the order is established. Typically, constraints prevent this from
occurring, so the strain leads to a long-ranged, frustrating interaction
(which gives rise, e.g. to the tweed pattern observed both in conventional
martensites and, recently, in CMR materials). The relation of this physics to
the observed sensititivity remains an open problem.

\section{Conclusion}

This short note has attempted to outline a 'botany' of causes for the
(surprisingly widespread) phenomenon of 'non-intrinsic behavior' of, as I
prefer to put it, sensitivity of properties to perturbations. The CMR
materials were argued to exhibit an unexpected sensitivity phenomenon
characterized not by an enhanced linear response to disorder but by a low
threshold to a qualitatively different nonlinear response. I suggest that
refining and extending the classification scheme given here is an important
task of materials physics and, as shown by the familiar example of
semiconductors, may concievably lead to new and useful materials functionalities.

\textit{Acknowledgements:} The authors work in this area was supported by the
US National Science Foundation under the MRSEC program.

\end{document}